\documentclass[conference]{IEEEtran}
\usepackage{cite}
\usepackage{amsmath,amssymb,amsfonts}
\usepackage{algorithmic}
\usepackage{graphicx}
\usepackage{textcomp}
\usepackage{xcolor}
\usepackage{multirow}
\usepackage{array}

\def\BibTeX{{\rm B\kern-.05em{\sc i\kern-.025em b}\kern-.08em
    T\kern-.1667em\lower.7ex\hbox{E}\kern-.125emX}}
\begin{document}

\title{Developing Decentralised Resilience to Malicious Influence in Collective Perception Problem}

\author{\IEEEauthorblockN{Chris Wise, Aya Hussein, Heba El-Fiqi}
\IEEEauthorblockA{\textit{School of Engineering and Information Technology} \\
\textit{University of New South Wales}\\
Canberra, Australia \\
c.wise@student.unsw.edu.au, a.hussein@adfa.edu.au, h.el-fiqi@unsw.edu.au}}

\maketitle

\begin{abstract}
In collective decision-making, designing algorithms that use only local information to effect swarm-level behaviour is a non-trivial problem. We used machine learning techniques to teach swarm members to map their local perceptions of the environment to an optimal action. A curriculum inspired by Machine Education approaches was designed to facilitate this learning process and teach the members the skills required for optimal performance in the collective perception problem. We extended upon previous approaches by creating a curriculum that taught agents resilience to malicious influence. The experimental results show that well-designed rules-based algorithms can produce effective agents. When performing opinion fusion, we implemented decentralised resilience by having agents dynamically weight received opinion. We found a non-significant difference between constant and dynamic weights, suggesting that momentum-based opinion fusion is perhaps already a resilience mechanism.
\end{abstract}


\section{Introduction \label{sec:Introduction}}
Collective (swarm-based) decision-making is ``the phenomenon whereby a collective of agents makes a choice in a way that, once made, it is no longer attributable to any of the individual agents" \cite{valentini2017best}. Examples of swarm-based decision-making include:
\begin{enumerate}
    \item a swarm capable of deciding which gesture-based command a human is issuing \cite{giusti2012distributed}
    \item swarm members concluding the fastest action without explicitly measuring action execution time \cite{scheidler2015k}
    \item robot selection and command via human to swarm interaction \cite{couture2010selecting}
    \item optimal site selection \cite{valentini2015efficient}
    \item swarm leader election \cite{bounceur2017logo}.
\end{enumerate}
However, the above problems used domain-specific algorithms that exploit environmental features, limiting these algorithms' applicability to other domains \cite{valentini2015efficient}. The collective perception problem was designed to act as an abstract problem that develops and benchmarks generic domain-agnostic algorithms. In the collective perception problem, swarm members explore a grid-based environment while evaluating a feature's abundance. The swarm must collectively determine the abundance of the feature, with this feature abstracted to cell colour (black or white) \cite{valentini2016collective}. The problem is complex as agents with only local sensing capabilities must reach a collective consensus regarding the global state of the environment. The problem is difficult to solve efficiently and effectively in a fully distributed method. The problem is analogous to practical swarm problems, such as searching for precious metals, pollutants, or pest-damaged areas in a farm \cite{valentini2016collective, ebert2020bayes}.
\par
Previous collective perception algorithms, detailed later in Section \ref{Related_Work}, implement local rules designed to control swarm-level behaviour. Through Machine Learning (ML), we propose that swarm members learn to effectively select an optimal action based on their perception of the environment and task progress. Swarm members will also learn how to detect and resist negative influence. We propose using Machine Education (ME) to guide the process of identifying, teaching, and evaluating these skills. ME borrows from the field of human education and focuses on systematically designing a curriculum driven by learning goals \cite{leu2017machine}. Although still in its infancy, ME has shown potential for producing multi-skilled agents \cite{clayton2019machine, hussein2021machine}. To our knowledge, this is the first study to investigate swarm resilience taught through ME.
\par
Previous literature has articulated swarm robustness as a desirable trait \cite{csahin2004swarm}. Swarm robustness to noise and, at its extreme, malicious influence is ever more important as swarming solutions are applied to more potentially life-threatening real-world problems. Examples of highly consequential situations include search and rescue (SAR) \cite{arnold2018search, cardona2019robot}, surveillance \cite{recchiuto2016visual}, space exploration \cite{truszkowski2004nasa} and nano-medicine \cite{al2012swarming}. Swarm robustness is linked to security issues being considered whilst developing swarm robotics. Suppose one was to wait for security issues to present once swarms are deployed in real-world applications. Waiting for deployment would likely result in time-consuming redesign or complete abandonment of an insecure approach. Potential attacks against swarms could include:
\begin{enumerate}
    \item tampering with agent sensors such that sent messages can contain wrong or deceptive information
    \item jamming communication channels
    \item capturing agents to destroy or degrade stored information \cite{higgins2009survey}.
\end{enumerate}
These attacks could have deadly consequences; for example, in a SAR situation, the swarm could be rendered incapable of locating a victim in time or wrongly concluding the victim's location.
\par
Previous work has attempted to implement swarm resilience, but most have implemented solutions in a non-fully decentralised way \cite{christensen2008fault, christensen2009fireflies, zikratov2016dynamic, saldana2017resilient}. Though, one fully decentralised solution using blockchain technology has been implemented \cite{strobel2018managing}. However, this blockchain approach assumes that agents have a robust local network connection capable of communicating complex data and have extensive computing hardware. Following the recommendations of \cite{hussein2021machine}, our proposed resilience approach minimises the communication bandwidth required between agents and assumes minimal hardware requirements. \par
This paper contributes:
\begin{enumerate}
    \item further investigations into the efficacy of ME when designing algorithms for complex multi-skilled swarm tasks; investigations needed as ME is still in a period of infancy
    \item a fully decentralised swarm resilience algorithm that builds upon previous approaches
    \item an abstracted simulation built to be more applicable than previous simulations to enable algorithms more agnostic to the physical implementation of agents 
    \item recommendations for future research into the collective perception problem and negative influence in swarms.
\end{enumerate}
\par
Following this, Section \ref{Related_Work} presents related work, starting with a further description of the collective perception problem and existing algorithms. Further background on existing swarm resilience approaches is also provided. Section \ref{Methodology} presents the proposed approach for teaching swarm members. The proposed approach is then evaluated in Section \ref{Results}. Considerations and recommendations based on experiment results are discussed in Section \ref{Discussion}. In Section \ref{Broader_Impact} impacts of the research are considered. Finally, conclusions are drawn, and directions for future work are provided in Sections \ref{Conclusions} \& \ref{Future_Work}, respectively.
\section{Related Work \label{Related_Work}}
The collective perception problem proposed by \cite{valentini2016collective} serves to develop and evaluate domain-agnostic swarming algorithms and has been investigated in several studies \cite{ebert2018multi, strobel2018managing, bartashevich2019benchmarking, hussein2020swarm, hussein2021stable, hussein2021machine}. In the collective perception problem, a wall-bounded grid-based environment contains cells with or without a feature $f$. This feature $f$ is represented by cell colour, with white cells containing $f$ and black cells not containing $f$. A swarm is then deployed in the environment and tasked to determine the abundance of $f$.
\newline 
Swarm members can:
\begin{enumerate}
    \item only sense the colour of the cell they occupy  
    \item only communicate (i.e. send and receive messages) with agents within the communication range $R_{comm}$
    \item excursively sense cell colour or broadcast their opinion; agents cannot perform both actions at the same time
    \item always receive opinions sent by other agents within $R_{comm}$ 
    \item not calculate their absolute position within the environment
    \item only move forward and turn left or right.
\end{enumerate}
\par
Problem complexity is affected by the ratio and spatial distribution of the feature $f$. The problem becomes more challenging as the feature ratio $r$ approaches 0.5 and becomes more accessible as $r$ tends to 0 or 1. When discussing the spatial distribution of $f$, a distinction is drawn between ``clustered" and ``non-clustered" environments. Clustered environments are more complex than non-clustered ones \cite{ebert2018multi, bartashevich2019benchmarking}. Non-clustered environments have cell colour uniformly distributed across the environment, and clustered environments split the grid into two same-coloured cell regions.
\par
Existing algorithms for the collective perception problem include the following behaviours:
\begin{enumerate}
    \item feature examination in the form of environment exploration and tile colour sensing
    \item opinion exchange being used to leverage the exploration performed by other agents. 
\end{enumerate}
These existing algorithms differ in the timing, duration, and selection of the above behaviours. Examples of previous implementations include swarm members that:
\begin{enumerate}
    \item alternate between exploration and opinion dissemination. Exploration duration is drawn from an exponential distribution. Dissemination duration is proportional to an agent's confidence, where highly confident agents disseminate their opinions for longer \cite{valentini2016collective}
    \item alternate between exploration and opinion dissemination. Opinion dissemination duration is proportional to an agent's confidence. However, exploration duration is the same for all agents \cite{ebert2018multi}.
\end{enumerate} 
In the above examples, agent confidence was based on the frequency of white cells in the agent's sample of tiles sensed. However, this method for setting agent confidence is sub-optimal and is biased to environments with a high $f$ ratio \cite{hussein2020swarm, hussein2021machine}.
\par
Examples of more recent and sophisticated algorithms include: 
\begin{enumerate}
    \item agents that maintain intrinsic opinion preferences and a human supervisor manipulating agent preferences to improve performance \cite{bartashevich2019ising}
    \item a fuzzy inference system (FIS) that sets an agent's confidence given its local observations \cite{hussein2020swarm}
    \item Bayes Bots (BB) that sense and broadcast for a constant duration. The bots then use a Bayesian model to determine when to finalise their decision \cite{ebert2020bayes}
    \item a confidence independent algorithm where all agents spent the same time examining the environment. In the opinion dissemination phase, a shepherding agent gathers swarm members together, enabling more efficient opinion fusion. Agents finalised their opinion when the calculated collective opinion reached a predefined threshold \cite{hussein2021stable}.
\end{enumerate}
State-of-the-art results have been achieved using reinforcement learning (RL) and machine learning (ML) techniques to map an agent's perception to an optimal action. Combined with RL and ML techniques, ME was used to systematically design a curriculum to teach and optimise agent performance \cite{hussein2021machine}. The curriculum optimised navigation, selection between broadcasting and sensing, and opinion finalisation. Agents developed by \cite{hussein2021machine} showed robustness to changes in swarm size and sensing and communication noise. However, the swarm was susceptible to malicious agents disrupting task success. This state-of-the-art could be extended by leveraging more recent RL algorithms such as DQN \cite{mnih2013playing}, A2C \cite{mnih2016asynchronous}, or PPO \cite{schulman2017proximal}, which have all shown performance improvements over the Q-Learning Algorithm used by \cite{hussein2021machine}.
\par
We define malicious members as those members who appear normal to regular members but act to disrupt task performance by slowing regular members or disrupting and degrading information propagation. Examples of malicious members acting to disrupt tasks include malicious members:
\begin{enumerate}
    \item moving to slow regular members finding land mines \cite{sargeant2013modelling}
    \item jamming communications between regular members to slow and degrade the collective consensus process \cite{sargeant2013modelling}
    \item intentionally reporting incorrect information and ``spoofing" the ID of regular members to influence the collective opinion to the wrong location of a target \cite{gil2017guaranteeing}
    \item intercepting regular agents and degrading their ability to perform the task \cite{zikratov2016dynamic}
    \item observing the environment and spreading information opposite to what has been observed \cite{guerrero2017formations}.
\end{enumerate}
\par
Specifically, for the collective perception problem, we define malicious members (malicious or negative influence) as those who broadcast an opinion opposite to the correct one. Without loss of generality, if the correct opinion were $a$, a malicious member would broadcast $b$. One such solution to malicious members in the collective perception problem used blockchain technology to identify and exclude the opinion of these Byzantine (malicious) Bots \cite{strobel2018managing}. The researchers used Ethereum-based smart contracts and had blocks mined by agents. The contract acted to distribute knowledge, record votes, and apply decision-making strategies. The smart contract implemented three functions: \textit{registerRobot}, \textit{applyStrategy}, and \textit{vote}. Experiments started with swarm members connecting to an auxiliary node; this auxiliary node initialised a smart contract. Members then sent the auxiliary node a transaction to the \textit{registerRobot}, and once all agents had registered, they disconnected from the auxiliary node, and the experiment started. 
\par
Agents ended their dissemination phase by connecting to the Ethereum processes of their local neighbours and mining received transactions. During this dissemination phase, members sent their transactions to the \textit{vote} function. This \textit{vote} function included the member's current opinion (and agent confidence when using DC), the number of blocks their opinion is based on, and the block hashes that the agent had received when listening to the events created by \textit{registerRobot} and \textit{applyStrategy}. An \textit{applyStrategy} function was then applied to votes. As in \cite{valentini2016collective}, three decision-making strategies were applied: 
\begin{enumerate}
    \item DMVD (voter model): adopt the opinion of a random neighbour
    \item DMMD (majority voting): adopt the opinion of the majority of the neighbours
    \item DC (direct comparison): adopt the opinion of a random neighbour if the neighbour's confidence is higher than yours.
\end{enumerate}
The \textit{applyStrategy} function then applied the decision-making strategy to only those opinions from stable blocks (a block with six or more confirmations). While members waited for their \textit{applyStrategy} transaction to be mined, agents neither disseminated their opinion nor explored the environment. However, agents resumed exploring the environment as soon as their transaction was mined.
\par
Three safety criteria were implemented into smart contracts to create swarm resilience. The smart contract would exclude an opinion if any safety criteria were violated. The three criteria checked if:
\begin{enumerate}
    \item the opinion of an agent was based on an outdated block number, with an opinion being considered outdated if the member had not sent a transaction to \textit{applyStrategy} in the last 25 blocks
    \item an agent had exhausted its assigned votes. Limiting vote numbers prevented ``vote spamming"
    \item an agent's opinion was based on a significantly different blockchain. By comparing hash values of blocks, the smart contract evaluated if an agent's opinion was aligned with the current consensus.
\end{enumerate}
\par
Researchers found that in non-malicious environments, the blockchain approach and classical approach \cite{valentini2016collective} achieved similar task performance \cite{strobel2018managing}. However, the blockchain approach was more resilient and likely to succeed in hostile environments than the classical approach \cite{strobel2018managing}. Despite its success, the blockchain approach slows consensus as communication is much slower and differing local sub-swarm opinions often manifest. This blockchain approach does not suit the usually limited hardware available to robots used for swarming solutions. Hardware limits are an issue as members must ``mine" transactions, which is computationally expensive and memory exhaustive. Limited hardware could ``mine" these transactions. However, the limited resources would result in slow mining times, slowing task performance. Finally, message size is 40 times larger in the blockchain approach compared to messages used in the classic approach of \cite{strobel2018managing}. This significant message size thus requires a larger bandwidth connection between agents.
\section{Methodology\label{Methodology}}
We propose using RL to teach agents the skills required for optimal performance in the collective perception problem. Like \cite{hussein2021machine}, the Dick and Carey Model \cite{dick2005systematic} will guide the design and implementation of our curriculum. Following this section is the application of the nine steps of the Dick and Carey Model to our task.
\newline
\subsection{Identify Instructional Goals}
The instructional goal is to enable swarm members to efficiently and effectively solve the collective perception problem in a distributed and autonomous manner. In a minimum amount of time, we require the swarm to collectively decide with minimal errors if the environment contains a majority of white or black tiles, that is, if $r \leq 0.5$ or $r > 0.5$, respectively.
\newline
\subsection{Instructional Analysis}
Given the strong results achieved by \cite{hussein2021machine}, we will use the same instructional analysis used in this previous work. However, we extend this instructional analysis by identifying a different skill designed to implement swarm resilience.
\par
We identify that swarm members require the ability to navigate (Skill 1.0). Though the environment does not contain obstacles, swarm members must learn to avoid the environment bounding walls and other swarm members. The navigation skill entails two sub-skills, navigation in unpopulated (Skill 1.1) and populated (Skill 1.2) environments. The second skill (Skill 2.0) requires swarm members to learn to select between sensing or broadcasting their opinion. The former behaviour is necessary to obtain correct opinions, while the latter is needed for reaching collective consensus. Swarm members will also learn when to commit to an opinion (Skill 3.0). Finally, extending beyond previous work, agents will learn to dynamically adjust the weighting given to opinions $a$ \& $b$ (Skill 4.0) when calculating the collective opinion (quorum sensing variable). Here, $a$ and $b$ represent the opinions corresponding to white and black tiles, respectively. This opinion weighting refers to the weight given to the received opinions when agents perform opinion fusion using a momentum-based method. Opinion fusion is calculated following Equation \ref{eqn:collectiveOpinionCalculation}.
    \begin{equation} \label{eqn:collectiveOpinionCalculation}
        \gamma_{k+1} = w\gamma_{k} + (1 - w)\Omega_{j}
    \end{equation} 
where $\gamma$ is the quorum sensing variable, $w$ is the weight that determines the influence of a newly received opinion $\Omega_{j}$. $w$ is set to 0.9 such that $\gamma_{k}$ is a robust representation of all the received opinions \cite{hussein2020swarm}.   
\newline
\subsection{Analysis of learners and Learning and Evaluation Context}
Next, we determine the characteristics of the following components: learners (swarm members), instructional environment (simulation) and evaluation environment (simulation). Essential characteristics of swarm members are listed in Section \ref{Related_Work}. Swarm members can also mutually only step forward or turn $90^\circ$ left or right for each navigation cycle. $R_{comm}$ is set so agents could only communicate at maximum with eight other agents in the immediate surrounding tiles. 
\par
The instructional and evaluation environments use the same simulation but differ in parameters. These differing parameters are detailed in Table \ref{parameterTable}. The simulation uses Python 3, and no physics has been simulated to minimise assumptions regarding the physical implementation of agents and the environment they may operate in (water, land, air). Thus, agents will not collide with each other or walls, and collisions are avoided as agents cannot move forward if an agent or wall is in front of them. Each iteration of an agent decision cycle (i.e. choosing to commit to an opinion, deciding to sense or broadcast, performing a navigation action) was simulated to take one second. Our simulation is more discrete when compared to previous simulations. For example, when agents move forward onto another tile, at no point are agents simulated between tiles; agents either occupy or do not occupy a given tile. Agents cannot move diagonally either; they always perform $90^{\circ}$ turns. We assume perfect sensing and communication. The simulation is available here: 
\textcolor{blue}{https://github.com/ChrisWise07/ZEIT3190-Swarm-Research}
\par
During learning, the ``teacher" will know the correct colour opinion of the environment (i.e. is $r \leq 0.5$ or $r > 0.5$). The ``teacher" also knows agent states (i.e. an agent's absolute position, current opinion, whether they are sensing or broadcasting, whether they have reached a final decision, and if that decision is correct). The ``teacher" facilitates the learning process by observing agents and providing feedback following the reward functions detailed in Subsection \ref{instructionStrategy}.To assess swarm performance in evaluation environments, the ``teacher" will still have the same knowledge and will use this knowledge to evaluate swarm performance. In evaluation, swarm members act autonomously to mimic real-world deployment.
\newline
\subsection{Defining Performance Measures}
The performance measures (PMs) for: 
\begin{enumerate}
    \item Skills 1.1 \& 1.2 are averaged across agents, and they are the average number of distinct cells visited by an agent per minute (PM 1.1) and the average total number of distinct cells visited by an agent at the end of an episode (PM 1.2)
    \item Skill 2.0 are the ratios of correct and incorrect opinions broadcast (PM 2.1 \& PM 2.2) and overall broadcasting accuracy (PM 2.3). How broadcasting accuracy is calculated can be found in Subsection \ref{broadcastSenseEvals}.
    \item Skill 3.0 are the average number of correct commitments (PM 3.2) and average decision time (PM 3.1)
    \item Skill 4.0 are the average distance an agent's quorum sensing variable is from the correct opinion (PM 4.1) and the average distance between the weighting an agent applies to opinions and the optimal opinion weightings (PM 4.2). PM 4.2 is calculated according to Equation \ref{optimalWeightingDistance}.
    \begin{equation} \label{optimalWeightingDistance}
        |w_{max} - w_{co} + w_{ic}|
    \end{equation}
    $w_{max}$ is the max weighting that can be given to an opinion, $w_{co}$ is the weighting given to the correct opinion and $w_{ic}$ is the weighting given to the incorrect opinion.
\newline
\end{enumerate}
\subsection{Developing Assessment Instruments}
Next, we develop a set of assessments to evaluate PMs. To assess:
\begin{enumerate}
    \item Skill 1.1, a single swarm member is deployed in the environment, and PM 1.1 \& 1.2 is measured for that single agent. The Skill 1.2 assessment is similar to Skill 1.1 assessment, except all swarm members are deployed in the environment, and PMs 1.1 \& 1.2 are measured and averaged across all agents
    \item Skill 2.0, all swarm members are deployed, and PMs 2.1 \& 2.2 \& 2.3 are measured across all agents. Learning Skill 1.0 is a prerequisite for performing Skill 2.0. Therefore, during the Skill 2.0 assessment, agents must navigate the environment while sensing or broadcasting
    \item Skill 3.0, the assessment is identical to the Skill 2.0 assessment, except agents can make a final decision before the maximum task time $T_{max}$ 
    \item Skill 4.0, the assessment is identical to the Skill 2.0 assessment, but agents can change the weightings allocated to opinions $a$ and $b$ when performing opinion fusion following Equation \ref{eqn:collectiveOpinionCalculation}.
\newline
\end{enumerate}
\subsection{Developing Instructional Strategy}\label{instructionStrategy}
As RL instructional strategies have been shown to effectively teach the identified skills \cite{hussein2021machine}, we will also use RL instructional strategies. We use Deep Q-Learning (DQN) \cite{mnih2013playing} for Skills 1.0 \& 2.0 \& 3.0 and Proximal Policy Optimization (PPO) \cite{schulman2017proximal} for Skill 4.0. We use the DQN and PPO algorithms implemented in Stable Baselines 3 \cite{raffin2019stable}.
\par
RL for Skill 1.0 uses the following states $S^{1.0} = {s_{k}^{1.0} = (o_{k}^{1.0}, p_{k}^{1.0})}$ such that $o_{k}^{1.0} = \{\textrm{none, } \textrm{wall, } \textrm{corner}, \textrm{agent}\}$ is the type of object directly in front of the agent, and $p_{k}^{1.0} = \{\textrm{front, } \textrm{left, } \textrm{right, } \textrm{frontLeft, } \textrm{frontRight}\}$ is the position of that object relative to the agent. Possible navigation actions are $A^{1.0} = \{\textrm{step forward, } \textrm{turn left, } \textrm{turn right}\}$. The reward $R^{1.0}$ urges agents to visit new cells, and this reward is calculated following Equation \ref{eqn:navReward}.
\begin{equation}\label{eqn:navReward}
   R^{1.0} = \begin{cases} 
      1 \textrm{ if } c \notin l, \\
      0 \textrm{ otherwise} \\
   \end{cases}
\end{equation}
such that $c$ is the cell the agent is currently in and $l$ is a set of unique cells the agent has visited. Note that the superscript 1.0 is used for components common to Skills 1.1 \& 1.2.
\par
RL for Skill 2.0 uses the following states $S^{2.0} = {s_{k}^{2.0} = (r_{k}^{2.0}, \Omega_{k}^{2.0}, \gamma_k^{2.0}, d_{k}^{2.0})}$ where the state variables are calculated as follows:
\begin{itemize}
    \item $r_{k}$ is a ratio representing the number of tiles sensed by an agent to the total number of tiles in the environment (the product of environment length $L$ and width $W$). It is calculated according to Equation \ref{eqn:ratioObserved}
        \begin{equation} \label{eqn:ratioObserved}
            r_{k}^{2.0} = \frac{\textit{number of observations}}{L \times W}
        \end{equation} 
    \item $\Omega_{k}$ is the agent's current opinion is calculated according to Equation \ref{eqn:agentOpinion}
        \begin{equation} \label{eqn:agentOpinion}
            \Omega_{k} = round(\frac{\textit{number of white cells observed}}{\textit{number of observations}})
        \end{equation} 
    \item $\gamma_{k}$ is the agent's quorum sensing variable is calculated following Equation \ref{eqn:collectiveOpinionCalculation} 
    \item $d_{k}$ is the difference between $\Omega_{k}$ and  $\gamma_{k}$ and is according to Equation \ref{eqn:differenceOfOpinion}. 
        \begin{equation} \label{eqn:differenceOfOpinion}
            d_{k} = |\Omega_{k} - \gamma_{k}|
        \end{equation} 
\end{itemize}
The possible actions for Skill 2.0 are $A^{2.0} = \{\textrm{sense, } \textrm{broadcast}\}$. The reward function rewards agents for sharing correct opinions and penalises agents for sharing incorrect opinions. Sensing incurs a slight penalty to discourage agents from sensing too much. $R^{2.0}$ is set according to Equation \ref{eqn:senseReward}.

\begin{equation}\begin{aligned}\label{eqn:senseReward}
R^{2.0} = \begin{cases}
      -2 \textrm{\phantom{-----}if broadcasting an incorrect opinion}, \\
      1 \textrm{\phantom{-------}if broadcasting a correct opinion}, \\
      -0.01 \textrm{\phantom{-}otherwise} \\
   \end{cases}
\end{aligned}
\end{equation}
\par
RL for Skill 3.0 uses the same set of states as those used for Skill 2.0. For Skill 3.0, there are two possible actions $A^{3.0} = \{\textrm{not commit, } \textrm{commit}\}$. The \textit{commit} action results in an agent making a final decision (i.e. opinion = Round($\gamma$)) and following this by navigating and always broadcasting Round($\gamma$). The \textit{do not commit} action causes the agent to continue to navigate whilst choosing to sense or broadcast. The reward for Skill 3.0 is formulated according to Equation \ref{eqn:initialCommitReward}.
\begin{equation}\label{eqn:initialCommitReward}
   R^{3.0} = \begin{cases} 
      100 \textrm{\phantom{---0.}if committed to correct opinion}, \\
      -400 \textrm{\phantom{-0.}if committed to wrong opinion}, \\
      -0.025 \textrm{\phantom{-}otherwise} \\
   \end{cases}
\end{equation}
\par
RL for Skill 4.0 uses the states $S^{4.0} = {s_{k}^{4.0} = (\Omega_{k}^{4.0}, \gamma_{k}^{4.0})}$, where $\Omega_{k}^{4.0}$ and $\gamma_{k}^{4.0}$ are the same as described for Skill 2.0 and are calculated according to Equations \ref{eqn:agentOpinion} \& \ref{eqn:collectiveOpinionCalculation}, respectively. For Skill 4.0, the possible action is a list containing two entries, representing the weighting for opinions $a$ \& $b$ respectively, that is $A^{4.0} =  [w_{a} \in [0.0 - w_{max}], w_{b} \in [0.0 - w_{max}]]$, where $w_{max}$ is the max opinion weighting. The reward for Skill 4.0 is formulated to encourage a high weighting for the correct opinion and a low weighting for the incorrect opinion. It is calculated following Equation \ref{eqn:weightingReward}.
\begin{equation}\label{eqn:weightingReward}
   R^{4.0} = \frac{1}{w_{max} - w_{co} + \epsilon} + \frac{1}{w_{io} + \epsilon} 
\end{equation}
where $w_{max}$ is the max opinion weighting (set at 0.1 as discussed above, with differing opinion weightings discussed further in Section \ref{Results}), $w_{co}$ is the weighting given the correct opinion, $w_{io}$ is the weighting given to the incorrect opinion, and $\epsilon$ is used to mitigate division by 0. 
\newline
\subsection{Developing and Selecting Instructional Materials}
We next consider and develop lesson content and sequencing. Given a RL paradigm, content is defined as the experiences swarm members encounter during training. It is important to use diverse experiences to ensure the applicability of learnt skills to various scenarios. The rest of this section considers the selection of experiences for Skills 1-4.
\newline
\subsubsection{Skill 1.1 \& 1.2 Instructional Material}
To align with previous literature we consider only obstacle-free environments \cite{valentini2016collective, ebert2018multi, hussein2020swarm, hussein2021stable, hussein2021stable}. However, we depart from previous literature and do not allow collisions between walls or agents to enable a simulation more agnostic to physical agent implementation. The navigational skills require swarm members only to visit different cells. Therefore, environment type and feature ratio are irrelevant. The only prerequisite for the navigation skill is that agents can perform basic motion skills (i.e. move forward, rotate right, and rotate left), which are built-in capabilities. Therefore, we can start teaching the navigation skill straight away. The ability to move in a free region is a prerequisite to teaching agents to move in populated environments; hence, Skill 1.1 will be taught before Skill 1.2.
\newline
\subsubsection{Skill 2.0 \& 3.0 \& 4.0 Instructional Material}\label{skill234LessonMaterial}
\begin{figure}[htbp]
\centering
    \includegraphics[width=\columnwidth]{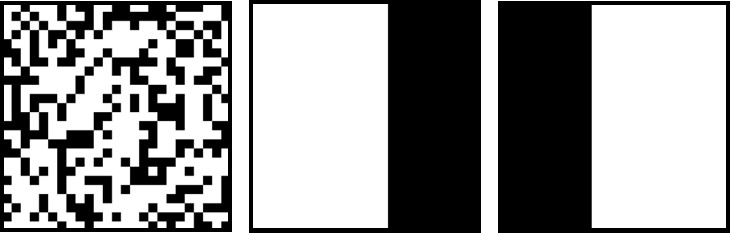}
    \caption{The three types of feature spatial distributions (r = 0.6): non-clustered (left), clustered with initial observations leading correct opinions (middle) and clustered with initial observations leading to incorrect opinions (right)}
\label{differentEvironmentTypes}
\end{figure}

As discussed in Section \ref{Related_Work}, two factors affect problem complexity: feature ratio and spatial distribution of tiles. Figure \ref{differentEvironmentTypes} illustrates the three types of feature spatial distributions we use: non-clustered, clustered with initial observations leading to the correct opinion and clustered with initial observations leading to the wrong opinion. Teaching Skills 3.0 \& 4.0 is irrelevant unless Skill 2.0 is adequately learnt as Skill 2.0 enables swarm members to form correct perceptions of the environment; these perceptions are used to determine decision finalisation (Skill 3.0) and dynamic opinion weightings (Skill 4.0). As curriculum complexity should increase over time \cite{bengio2009curriculum}, we implement PMs to indicate lesson complexity and set the probability of a feature spatial distribution type being used for subsequent lessons inversely proportional to swarm performance in the current learning iteration. Initially, all feature spatial distributions are equally likely. We use a truncated normal distribution \cite{burkardt2014truncated}, centred at 0.5 and truncated at 0 and 1 to set a lesson's feature ratio.
\newline
\subsection{Formative Evaluation and Instruction Revision}
To enable swift evaluations, we assess the instructions for each skill in miniature environments, allowing isolated testing of each skill module. The parameters of these miniature environments are listed in Table \ref{parameterTable}. The formative evaluation results are presented in Subsection \ref{Formative Evaluations}.

\begin{table}[htbp]
\caption{The parameter settings used for instructional and formative and summative evaluation environments. $u$ is unit length and is such that the dimensions of a cell are $1\times1$ $u^{2}$. $L$ and $W$ refer to environment dimensions, $N$ is the number of swarm agents, and $T_{max}$ is the maximum episode time}
\label{parameterTable}
\begin{tabular}{p{.2\columnwidth} p{.2\columnwidth} p{.2\columnwidth} p{.2\columnwidth}} 
    \hline
    \multicolumn{4}{l}{Formative evaluation parameters}\\
    L & W & N & $T_{max}$ \\
    38 & 38 & 25 & 125 \\
    \hline
    \multicolumn{4}{l}{Learning parameters} \\
    L & W & N & $T_{max}$ \\
    75 & 75 & 50 & 250 \\
    \hline
    \multicolumn{4}{l}{Summative evaluation parameters} \\
    L & W & N & $T_{max}$ \\
    150 & 150 & 100 & 500 \\
    \hline
\end{tabular}
\end{table}

\section{Results \label{Results}}
\subsection{Formative Evaluations \label{Formative Evaluations}}
\subsubsection{Skill 1.1 \& 1.2 \label{navigationEvals}}
\begin{figure}[htbp]
    \includegraphics[width=\columnwidth]{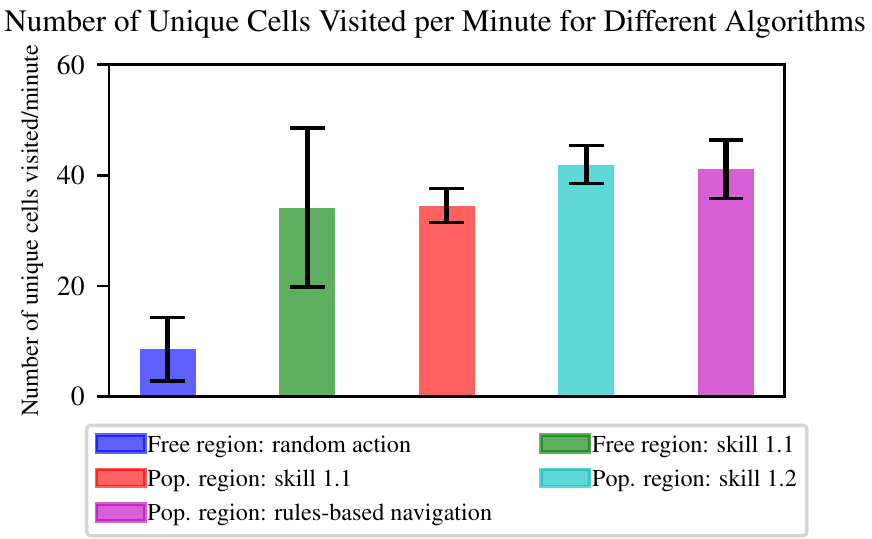}
    \caption{The number of unique cells visited per minute in free and populated regions before and after teaching the navigation skills. Column height represents the average value of PM 1.1 and the error bars represent the standard deviation, as calculated over 10 episodes of 5 minutes of navigation}
\label{cells_per_minute}
\end{figure}

\begin{figure}[htbp]
    \includegraphics[width=\columnwidth]{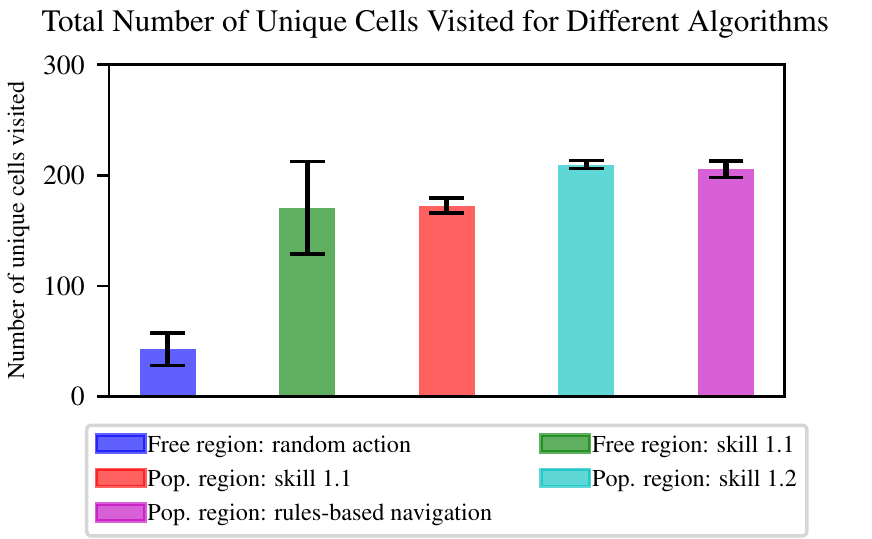}
    \caption{The total number of unique cells visited in an episode in free and populated regions before and after teaching the navigation skills. Column height represents the average value of PM 1.2 and the error bars represent the standard deviation, as calculated over 10 episodes at the end of 5 minutes of navigation}
\label{total_number_of_cells}
\end{figure}

Baseline performance for Skill 1.1 was evaluated by deploying a single agent in a free region environment. For this baseline, the agent randomly selected a navigation action (i.e. move forward, turn left, and turn right) at each time step. The evaluation consisted of ten 5-minute episodes where PM 1.1 was recorded every minute, and PM 1.2 was measured at the end of episodes. After teaching Skill 1.1 in an unpopulated environment, with all other parameters set as per Table \ref{parameterTable}, ten 5-minute episodes that measured PM 1.1 \& PM 1.2 were conducted. Figures \ref{cells_per_minute} \& \ref{total_number_of_cells} demonstrate the efficacy of Skill 1.1 curriculum as the agent increased the average rate of distinct cells visited per minute from 8.54 (std = 5.76) to 34.14 (std = 14.37) and total number of distinct cells from 42.7 (std = 14.86) to 170.7 (std = 41.7). PMs 1.1 \& 1.2 were deemed sufficient, and thus we moved to train Skill 1.2. To baseline Skill 1.2 performance, swarm members who learnt only Skill 1.1 were deployed in a populated environment. To further baseline Skill 1.2, we developed a rules-based navigation algorithm where agents:
\begin{enumerate}
    \item always moved forward if there was no object (agent, wall, corner) blocking them from doing so 
    \item randomly turned left or right if an agent or wall blocked the way forward
    \item always turned inward when facing a corner.
\end{enumerate}
Like earlier, ten 5-minute evaluations were conducted whilst measuring PMs 1.1 \& 1.2. Skill 1.2 was then taught according to lesson material discussed in Section \ref{Methodology} with parameters set as per Table \ref{parameterTable}. Figures \ref{cells_per_minute} \& \ref{total_number_of_cells} show a difference when navigating in populated regions for PMs 1.1 \& 1.2 for Skill 1.1 (mean = 34.53, std = 3.05) and (mean = 172.63, std = 6.82), respectively, and Skill 1.2 (mean = 41.92, std = 3.42) and (mean = 209.62, std = 3.71). However, there is little difference in PMs 1.1 \& 1.2 for Skill 1.2 and our rules-based navigation algorithm (mean = 41.08, std = 5.32) and (mean = 205.38, std = 7.62), respectively. Thus, we decided to instead use our rules-based navigation algorithm over Skill 1.2 as it is computationally less demanding. Therefore, our rules-based navigation algorithm would require less computing resources from a real-world agent.

\subsubsection{Skill 2.0 \label{broadcastSenseEvals}}
\begin{figure}[htbp]
\centering
    \includegraphics[width=\columnwidth]{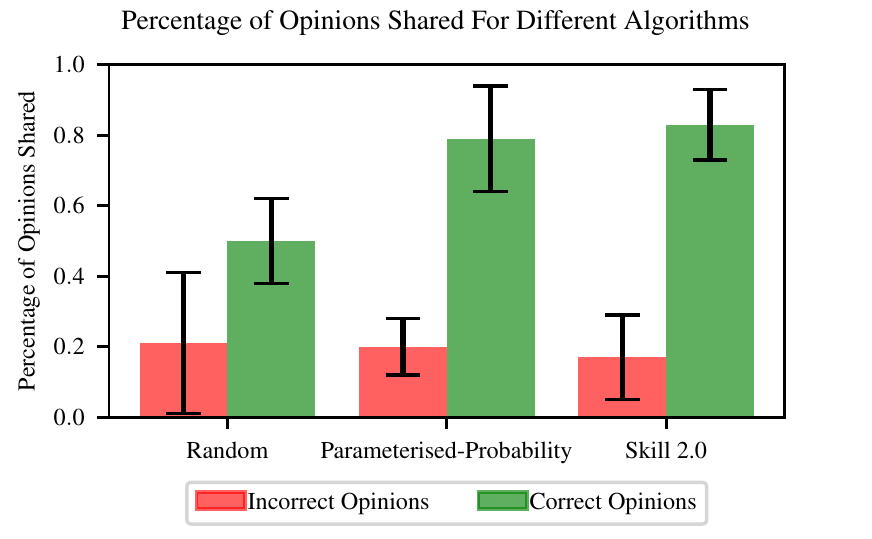}
    \caption{A comparison between the ratio of correct and incorrect opinions shared before and after teaching the broadcast sense skill. Column height represents the average value of PMs 2.1 and PM 2.2 and the error bars represent the standard deviation, as calculated from 90 experiments}
\label{correct_incorrect_shared}
\end{figure}

\begin{figure}[htbp]
\centering
    \includegraphics[width=\columnwidth]{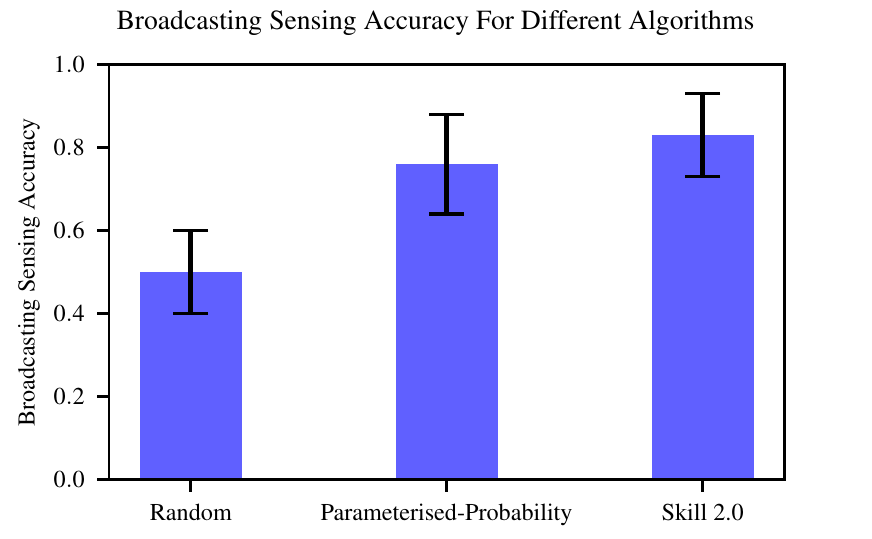}
    \caption{Broadcasting sensing accuracy before and after teaching the broadcast sense skill. Column height represents the average value of PM 2.3 and the error bars represent the standard deviation, as calculated from 90 experiments}
\label{broadcast_sensing_accuracy}
\end{figure}

Two prior tests were conducted to baseline Skill 2.0. In one baseline, agents would randomly choose between sensing and broadcasting. In the other baseline test, we parameterised the probability of an agent broadcasting. This parameterised probability averaged the percentage of tiles sensed by the agent and the inverse of the distance between the agent's opinion and quorum sensing variable. The parameterised probability was calculated according to Equation \ref{eqn:probabilityOfBroadcast}.
\begin{equation} \label{eqn:probabilityOfBroadcast}
    \frac{1}{2} [r_{k} + (1 - |\Omega_{k} - \gamma_{k}|)]
\end{equation}
Teaching Skill 2.0 used the parameters listed in Table \ref{parameterTable} and lesson sequencing was implemented as described in Subsection \ref{skill234LessonMaterial}. For all evaluations, ten instances of nine scenarios (i.e. 90 test scenarios) were used to evaluate the performance. The nine scenarios are the product of the feature ratios (0.52, 0.62, 0.72) applied to the three different tile spatial distributions. Figure \ref{correct_incorrect_shared} shows that the Skill 2.0 module was effective at producing the desired behaviour as improvements in PMs 2.1 and PM 2.2 were achieved (mean = 0.17, std = 0.12) and (mean = 0.83, std = 0.1), respectively. These results are compared to (mean = 0.21, std = 0.20) and (mean = 0.50, std = 0.12) for random choice and (mean = 0.20, std = 0.08) and (mean = 0.79, std = 0.15) for parameterised probability. We then extended upon previous work and measured the true positives, false positives, true negatives and false negatives and defined each as following:
\begin{enumerate}
    \item True positive (TP): an agent broadcasting the correct opinion
    \item False positive (FP): an agent broadcasting the incorrect opinion
    \item True negative (TN): an agent not broadcasting the incorrect opinion
    \item False negative (FN): an agent not broadcasting the correct opinion.
\end{enumerate}
With these statistical metrics, broadcasting sensing accuracy (PM 2.3) was measured according to Equation \ref{eqn:sensebroadcastAccuracy}.
\begin{equation} \label{eqn:sensebroadcastAccuracy}
    \frac{TP + TN}{TP + FP + TN + FN}
\end{equation}
Figure \ref{broadcast_sensing_accuracy} further demonstrates the effectiveness of the Skill 2.0 module as after training PM 2.3 was measured as (mean = 0.83, std = 0.1). This accuracy is compared to (mean = 0.5, std = 0.1) for random choice and (mean = 0.76, std = 0.14) for parameterised probability. We recommend measuring these generalised statistical metrics for future studies, as doing so one can provide sensitivity and specificity measures or aggregate metrics such as F1 scores or accuracy calculations \cite{fawcett2006introduction}.
\subsubsection{Skill 3.0 \label{commitEvals}}
\begin{figure}[htbp]
    \includegraphics[width=\columnwidth]{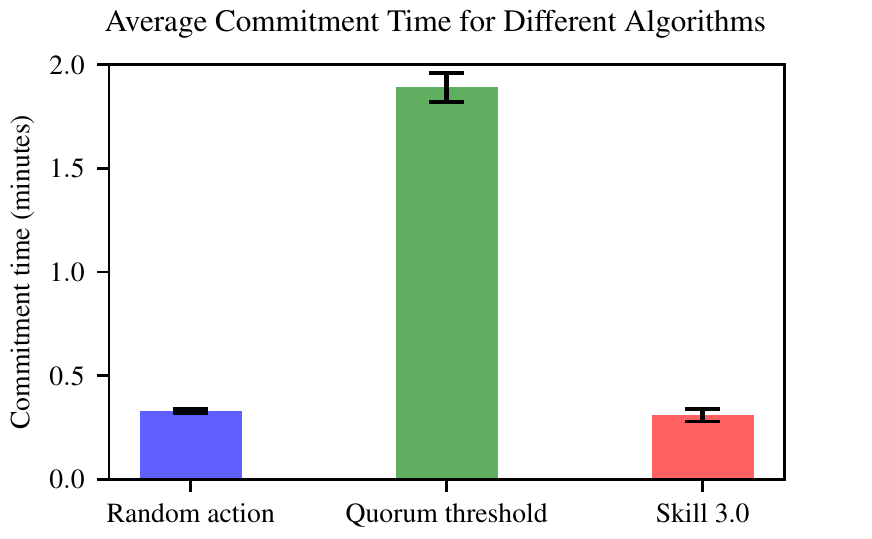}
    \caption{Performance measure PM 3.2 is shown when evaluating baselines and Skill 3.0. Column height represents the average time taken to commit (PM 3.2) and the error bars represent the standard deviation, as calculated from 90 experiments}
\label{commitTime}
\end{figure}

Skill 3.0 was taught similar to Skill 2.0, except while learning Skill 3.0, agents could reach a final decision. Before teaching Skill 3.0, two baseline evaluations were performed. For one baseline test, the agent would \textit{commit} with a probability of 5\% at each time step. For the other baseline test, until agents have observed more than one width of the environment, the agent could only navigate and sense tile colour. That is, until $r_{k} > \frac{1}{L}$, the agent could only sense and not broadcast. In this baseline test, agents would commit if their quorum sensing variable was sufficiently converged. That is, the quorum was within some limit, $\theta$, of 0 or 1 (i.e. $\gamma_{k} < \theta$ or $1 - \gamma_{k} < \theta$ ). Like Skill 2.0, 90 test scenarios were used, and tests would last for either $T_{max}$ minutes or until all agents had committed, whatever occurred first. Figure \ref{commitTime} shows that after learning Skill 3.0, agents would quickly commit, achieving a mean commitment time of 0.31 minutes (std = 0.03). However, this quick commitment came at a cost to accuracy, with, on average only 49.96\% (std = 3.71\%) of agents committing to the correct opinion. This quick commitment time and low accuracy suggested that the Skill 3.0 module should be redesigned. However, the quorum convergence method was extremely accurate, with an average of 98.27\% (std = 1.64\%) of agents committing to the correct opinion. This commitment accuracy came at little cost to task time, with agents committing on average in 1.89 minutes (std = 0.07). As the quorum convergence method was sufficiently accurate and fast, we did not redesign the Skill 3.0 curriculum as we deemed the quorum convergence method effective. Further, the quorum convergence method would require less computing resources from agents, and consequently, it is more applicable than Skill 3.0 to real-world agents.

\begin{figure}[htbp]
    \includegraphics[width=\columnwidth]{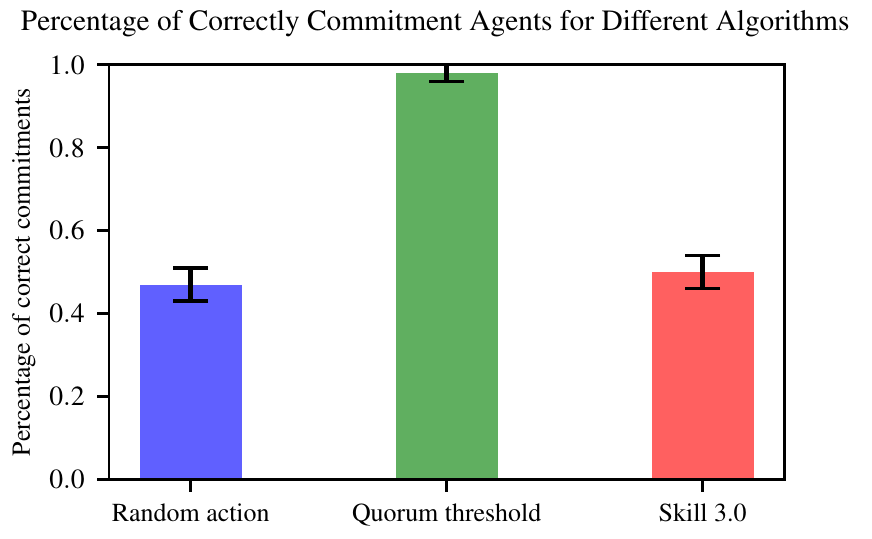}
    \caption{Performance measure PM 3.1 is shown when evaluating baselines and Skill 3.0. Column height represents the average commitment accuracy (PM 3.1) and the error bars represent the standard deviation, as calculated from 90 experiments}
\label{commitAccuracy}
\end{figure}
\newpage
\subsubsection{Skill 4.0 \label{weightingEvals}}
Skill 4.0 was taught similarly to Skill 2.0, except agents were allowed to set the weightings given to opinions $a$, $b$, at the beginning of each time step, and 5\% of swarm members were malicious. Three baseline evaluations, using the 90 test scenarios, were also conducted before teaching Skill 4.0. For one baseline test, agents navigated the environment while sensing or broadcasting; opinion weightings thus remained \textit{static}. For the other baseline test, an agent used an equation based on the distance between an opinion and the current quorum sensing variable to set the weighting at the beginning of each time step for opinions $a$ \& $b$, respectively. That is, without loss of generality, this \textit{equation-based} weighting gave more weighting to opinion $a$ and less to the historical quorum as the quorum sensing variable converged closer to $a$. This \textit{equation-based} weighting was calculated according to Equation \ref{weightingEquation}.
\begin{equation}\label{weightingEquation}
   \lambda(1 - |\Omega - \gamma|)
\end{equation}
where $\lambda$ is some constant controlling the max weighting given to new opinions. The \textit{equation-based} weighting was designed as a resilience mechanism. \textit{Equation-based} weighting acts such that as the collective opinion converges to the correct opinion, opinions opposite to the correct opinion (i.e. malicious opinions) will have less influence on the collective opinion. We also designed an \textit{inverted-opinion-based} weighting equation. This equation aimed to weigh the quorum more as it converged closer to an opinion as the quorum is thus more likely to be correct. That is, without loss of generality, this \textit{inverted-equation-based} weighting gave less weighting to opinion $a$ and more to the historical quorum as the quorum sensing variable converged closer to $a$. This \textit{inverted-equation-based} weighting was calculated according to Equation \ref{invertedWeightingEquation}.
\begin{equation}\label{invertedWeightingEquation}
   \lambda(1 - |(\Omega + 1)\bmod2 - \gamma|)
\end{equation}
where $\bmod2$ is being used as $\Omega \in \{0, 1\}$. 
\par
\begin{figure}[htbp]
    \includegraphics[width=\columnwidth]{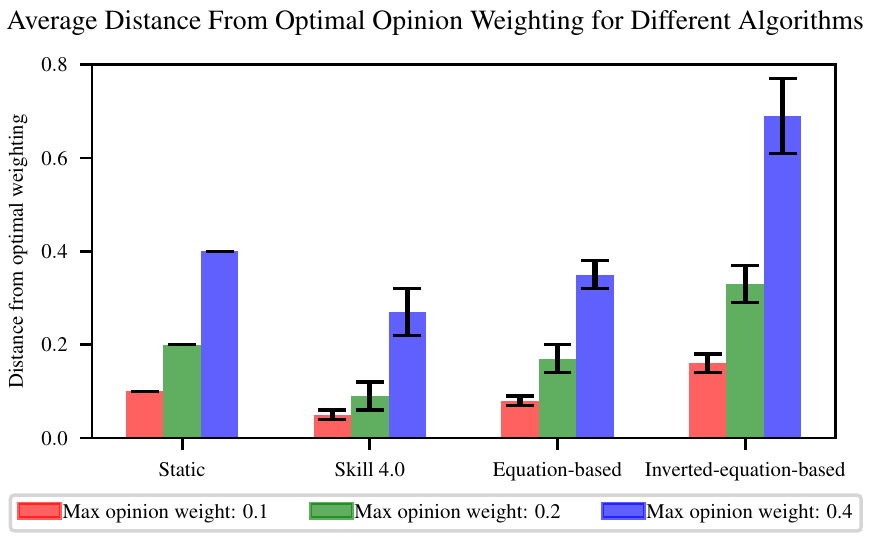}
    \caption{PM 4.2 when evaluating baselines and Skill 4.0. Column height represents the average distance between the optimal weighting and the actual weighting given to opinions (PM 4.1) and the error bars represent the standard deviation, as calculated from 90 experiments}
\label{optimalWeightingChart}
\end{figure}

We also measured the effect of varying the max weighting that can be given to an opinion (i.e. $w_{max}$). We used max weightings of (0.1, 0.2, 0.4). Following Equation \ref{optimalWeightingDistance} we calculated the max distance to be from an optimal weighting for a given a maximum weighting. Calculations are as follows:
\begin{enumerate}
    \item 0.1: $0.1 - 0 + 0.1 = 0.2$
    \item 0.2: $0.2 - 0 + 0.2 = 0.4$
    \item 0.4: $0.4 - 0 + 0.4 = 0.8$.
\end{enumerate}
Figure \ref{optimalWeightingChart} shows that the Skill 4.0 module was effective as agents during testing were applying weightings that maximised the influence of the correct opinion and minimised the influence of the incorrect opinion. PM 4.2 evaluated the efficacy of these weightings and PM 4.2 for a max weighting is 0.1 - (mean = 0.05, std = 0.01), 0.2 - (mean = 0.09, std = 0.03) and 0.4 - (mean = 0.27, std = 0.05). Aligning with this, the quorum sensing variable was sufficiently close to the correct opinion across the different max weightings: 0.1 - (mean = 0.24, std = 0.09), 0.2 - (mean = 0.23, std = 0.08) and 0.4 - (mean = 0.35, std = 0.07). Unsurprisingly, the \textit{inverted-equation-based} weighting resulted in the greatest distances from the optimal weightings. However, \textit{inverted-equation-based} weighting resulted in the lowest distances between the quorum and the correct opinion. As \textit{equation-based} and \textit{inverted-equation-based} methods performed sufficiently, like Skills 1.0 \& 3.0, we continued forward with the \textit{equation-based} and \textit{inverted-equation-based} weighting methods as they are computationally cheaper than Skill 4.0, again, making them more applicable in real-world situations. 
\par
To increase the efficacy of the \textit{equation-based} and \textit{inverted-equation-based} weighting methods, we had dynamic weightings set at every opinion update instead of once at the start of each time step. To accomplish this, when using the \textit{equation-based} weighting, we set the quorum sensing variable to update according to Equation \ref{dyanmicWeightingQuorumUpdate}.

\begin{equation}\label{dyanmicWeightingQuorumUpdate}
    \begin{split}
        \gamma_{n+1} = [1 - \lambda(1 - |\Omega - \gamma_{n}|)]\gamma_{n} \\ 
        + \lambda(1 - |\Omega - \gamma_{n}|)\Omega
    \end{split}
\end{equation}

Likewise, when using the \textit{inverted-equation-based} weighting we set the quorum sensing variable to update according to Equation \ref{dyanmicInvertedWeightingQuorumUpdate}.

\begin{equation}\label{dyanmicInvertedWeightingQuorumUpdate}
    \begin{split}
        \gamma_{n+1} = [1 - \lambda(1 - |(\Omega + 1)\bmod2 - \gamma_{n}|)]\gamma_{n} \\ 
        + \lambda(1 - |(\Omega + 1)\bmod2 - \gamma_{n}|)\Omega
    \end{split}
\end{equation}

\begin{figure}[htbp]
    \includegraphics[width=\columnwidth]{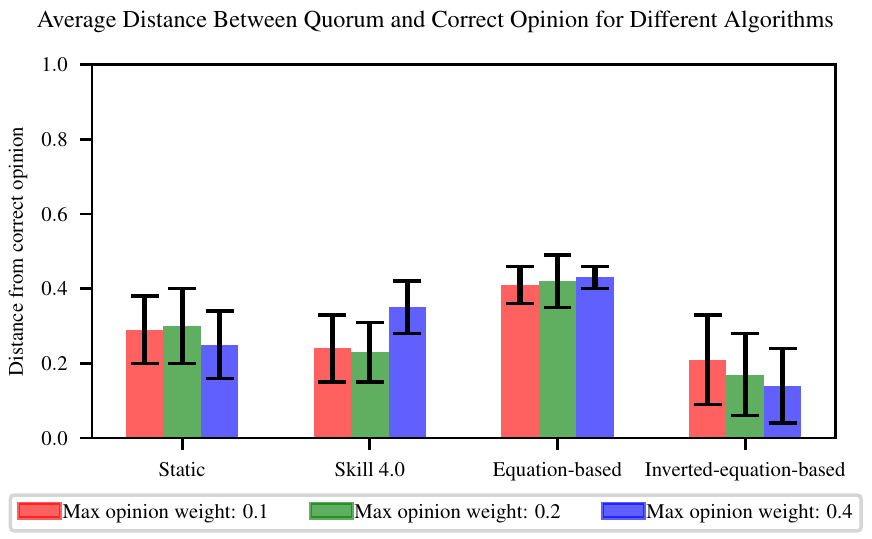}
    \caption{PM 4.1 when evaluating baselines and Skill 4.0. Column height represents the average distance between the quorum and the correct opinion (PM 4.1) and the error bars represent the standard deviation, as calculated from 90 experiments}
\label{collectiveOpinionDistanceChart}
\end{figure}

\subsection{Summative Evaluations \label{Summative Evaluations}}
The following section presents the summative evaluations after teaching Skills 1.0 - 4.0. As justified in Subsection \ref{Formative Evaluations}, agents used only Skill 2.0, and the remaining skills were replaced with the rule-based methods. The summative evaluations are similar to the formative evaluations except for changes in parameters as indicated in Table \ref{parameterTable}. Additionally, we changed the feature ratios to assess performance in scenarios not encountered during the teaching. The summative evaluations used the feature ratios 0.55, 0.65 \& 0.75. Further, when testing with malicious agents, 10\% of the swarm was malicious. In the first half of this section, we compared evaluations with and without malicious members, with and without dynamic weightings and across various max opinion weightings. In the second half of this section, we compared results against the state-of-the-art.
\newline
\subsubsection{Resilience to Malicious Influence}\label{Resilience to Malicious Influence}
When studying the effect of malicious influence on task performance, we used three different opinion weighting methods. These weighting methods, as detailed in Subsection \ref{weightingEvals}, where the \textit{static}, \textit{equation-based} \& \textit{inverted-equation-based}. We also studied the effect of the max opinion weighting using values 0.1, 0.4 \& 1.0. The opinion weighting of 1.0 was specifically used as it simplifies the \textit{equation-based} \& \textit{inverted-equation-based} methods. That is, with a max opinion weighting of 1.0, the \textit{equation-based} weighting was calculated according to Equation \ref{eqn:simpleEquationWeight}.    
\begin{equation}\label{eqn:simpleEquationWeight}
        \gamma_{n+1} = |\Omega - \gamma_{n}|\gamma_{n} + (1 - |\Omega - \gamma_{n}|)\Omega
\end{equation}
Accordingly, with a max opinion weighting of 1.0, \textit{inverted-equation-based} weighting was calculated according to Equation \ref{eqn:simpleInvertedEquationWeight}.    
\begin{equation}\label{eqn:simpleInvertedEquationWeight}
    \begin{split}
        \gamma_{n+1} = |(\Omega + 1)\bmod2 - \gamma_{n}|\gamma_{n} \\ + (1 - |(\Omega + 1)\bmod2 - \gamma_{n}|)\Omega
    \end{split}
\end{equation}
\par
\begin{figure}[htbp]
    \includegraphics[width=\columnwidth]{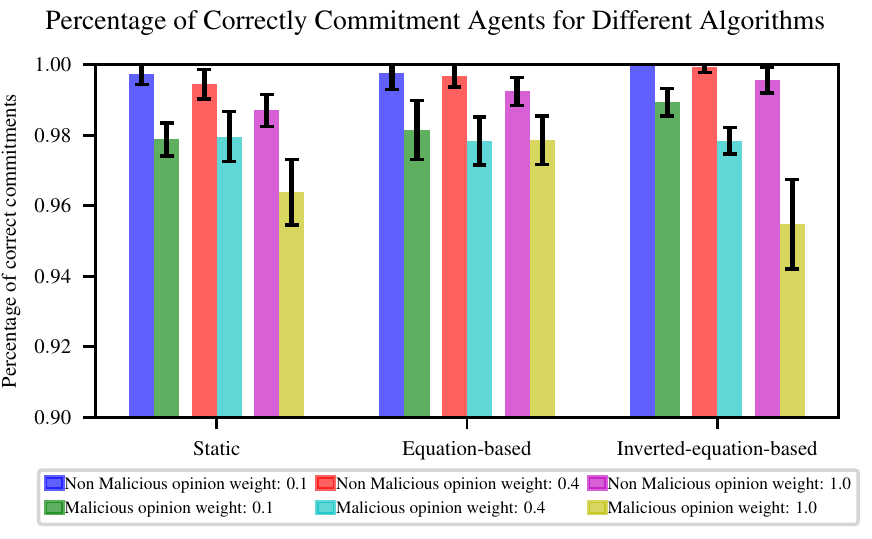}
    \caption{PM 3.1 when evaluating different weighting algorithms with and without malicious influence. Column height represents the average commitment accuracy (PM 3.1) and the error bars represent the standard deviation, as calculated from 90 experiments}
\label{accuracyFinalChart}
\end{figure}
\begin{figure}[htbp]
    \includegraphics[width=\columnwidth]{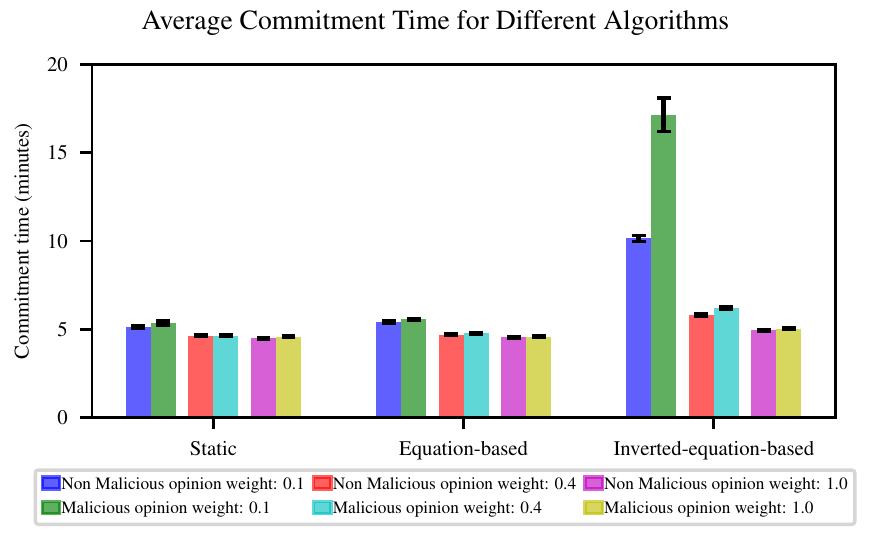}
    \caption{PM 3.2 when evaluating different weighting algorithms with and without malicious influence. Column height represents the average commitment time (PM 3.2) and the error bars represent the standard deviation, as calculated from 90 experiments}
\label{timeFinalChart}
\end{figure}

Figures \ref{accuracyFinalChart} \& \ref{timeFinalChart} show that task performance has a low sensitivity to a small amount of malicious influence. Further, as the results support, even without a dynamic weighting mechanism, the momentum-based opinion fusion enables the quorum to become a robust representation of received opinions \cite{hussein2020swarm}. However, with a dynamic weighting mechanism, the swarm does demonstrate increased resilience, most notably in reduced differences in task timings. That is, with \textit{equation-based} weightings, task time suffers the least with malicious members present. Increasing the max opinion weighting decreased task time but decreased commitment accuracy as well. We hypothesise that the increase in max opinion weighting enables a greater saturation of opinions, thus decreasing task time. However, a decrease in mean accuracy occurs with this higher opinion weighting as incorrect opinions subsequently have a greater influence.
\par
\begin{figure}[htbp]
    \includegraphics[width=\columnwidth]{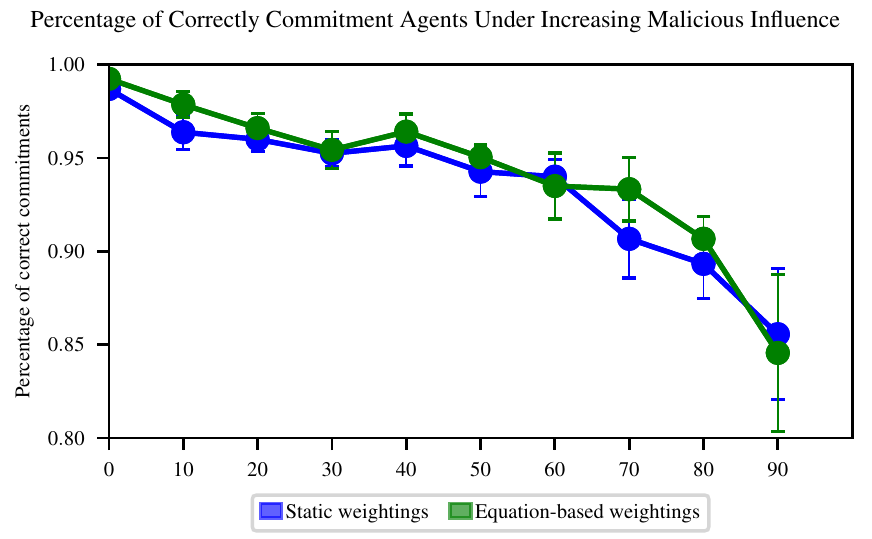}
    \caption{PM 3.1 when evaluating different weighting algorithms with and without malicious influence. Column height represents the average commitment accuracy (PM 3.1) and the error bars represent the standard deviation, as calculated from 90 experiments}
\label{accuracyFinalIncreasingMalChart}
\end{figure}
\begin{figure}[htbp]
    \includegraphics[width=\columnwidth]{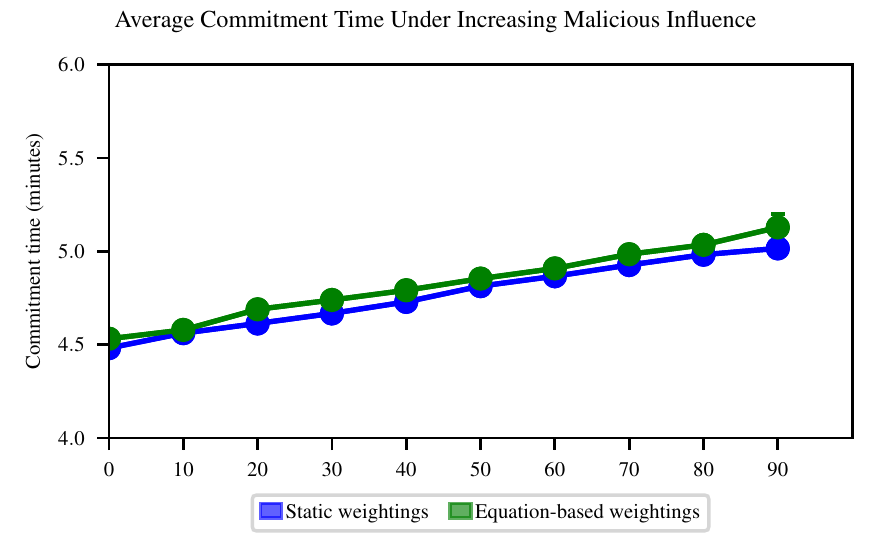}
    \caption{PM 3.2 when evaluating different weighting algorithms with and without malicious influence. Column height represents the average commitment time (PM 3.2) and the error bars represent the standard deviation, as calculated from 90 experiments}
\label{timeFinalIncreasingMalChart}
\end{figure}

As a max opinion weighting of 1.0 achieved that fastest task time, with little cost to accuracy, we decided to measure how resilient the swarm is to increasing numbers of malicious members using this configuration. We evaluated the \textit{equation-based} weighting method, using \textit{static} weightings as a control. As Figures \ref{accuracyFinalIncreasingMalChart} \& \ref{timeFinalIncreasingMalChart} show, increasing malicious influence decreases task performance. Accuracy decreases, and task time increases as more malicious members are introduced. Both the \textit{static} and \textit{equation-based} weighting methods showed resilience to this malicious influence. From 0\% to 90\% malicious influence, mean commit accuracy decreased by only 13.14\% \& 14.69\%, for \textit{static} and \textit{equation-based}, respectively. Likewise, from 0\% to 90\% malicious influence, mean task time increased by only 0.53 \& 0.59 minutes for \textit{static} and \textit{equation-based}, respectively. However, there is no statistical significance between the \textit{equation-based} and \textit{static} weighting methods. We calculated and averaged p-values across the malicious percentages and the average accuracy p-value = 0.56 and the average task time p-value = 0.71. We conjecture that there are two reasons for no significant difference between the two weighting methods. Firstly, the momentum-based opinion fusion is most likely already a resilience mechanism. Secondly, malicious agents and regular agents were initially spatially distributed in a clustered fashion; that is, malicious agents are next to other malicious agents, and regular agents are next to other regular agents. Since the agents are limited to local communication only, this limits the influence of incorrect opinions. Hence, the resiliency mechanism was not needed or necessary. Consequently, there is little difference between the opinion weighting methods. 

\newpage
\subsubsection{Comparison with other approaches}
This subsection presents a comparison between the performance of our proposed approach and seven other state-of-the-art approaches: Voter, Majority, DC \cite{valentini2016collective}, fuzzy confidence estimation (Fuzzy) \cite{hussein2020swarm}, Bayes Bots \cite{ebert2020bayes} (BB), shepherd-assisted (Shepherd) \cite{hussein2021stable} and the previous machine education approach (Previous-ME) \cite{hussein2021machine}. Note that we used data reported in \cite{hussein2021machine} for these comparisons as we matched the environment parameters used in the study. For a fair comparison, we have used the results we achieved using \textit{static} opinion weightings and a max opinion weighting of 0.1.
\newline
Our approach achieved a mean commitment time of approximately $\frac{1}{5}$ of the previous bests. Our approached achieved mean time of 5.11 (std = 0.07) minutes, compared to the previous bests, Previous-ME (mean = 23, std = 2.3 minutes) and BB (mean = 24.22, std = 17.18 minutes). For mean commitment accuracy (PM 3.2), our approach achieved accuracy as good as the previous best. Our approached achieved mean accuracy of 99.74\% (std = 0.31), compared to the previous bests, Fuzzy (mean = 99.43\% , std = 7.4) and Previous-ME (mean = 98.2\% , std = 10.74). Overall, our approach achieved accuracy as good as the previous approaches while achieving a task time significantly faster than any previous approach. However, these state-of-art approaches should be implemented within our simulation to enable a rigorous comparison.
\newline
Finally, we compared the swarm resilience of our approach against the Previous-ME approach. With our resilience mechanisms, we achieved less of a decrease in task performance compared to the Previous-ME approach. For this comparison, we used our model with \textit{equation-based}  weighting with a maximum opinion weighting of 1.0 and 5\% of the swarm being malicious. Table \ref{changeDueToMaliciousTable} details these comparisons. However, as discussed in Subsection \ref{Resilience to Malicious Influence}, in our experiments, malicious members may be limited in their influence. Thus we likely measured less of an impact from malicious members resulting from their limited influence, not resulting from our resilience mechanism.
\begin{table}[htbp]
\centering
\caption{Mean Commitment Time}
\begin{tabular}{|c|c|c|}
\cline{2-3}
\multicolumn{1}{c|}{} & Previous-ME & Our-Approach   \\
\hline
Change in Task Time (Minutes) & 1.37 & 0.05   \\
\hline
Change in Commitment Accuracy (\%) & 12.2 & 3.32   \\
\hline
\end{tabular}
\label{changeDueToMaliciousTable}
\end{table}

\section{Discussion \label{Discussion}}
The collective perception problem is challenging as agents must utilise local sensing and communication capabilities to reach a consensus on the global state of the environment. Previous approaches to the problem had agents applying rule-based algorithms designed by domain experts. This research applied ME to enable a holistic approach to deploying ML to agent behaviour selection. Overall, our ME approach to the collective perception problem has been successful as we have achieved performance better than the state-of-the-art. Specifically, we recorded the highest commitment accuracy and shortest task time and showed resilience to negative influence. 
\par
We have considerations for any future application of ME. The decomposition of behaviours into modules enables a curriculum to be designed easier, and sources of error can be identified more easily when failures occur. However, though the modular approach can result in high performance, one can not be sure that optimising the performance of each skill produces a decision-making strategy that optimises whole task performance  \cite{hussein2021machine}. However, a strategy to pseudo-optimise overall task performance is to change the PMs for each sub-skill. The PMs for all skills could become the mean percentage of correct commitments (PM 3.1) and task time (PM 3.2). For example, pseudo-whole-task-optimisation could be implemented to evaluate Skill 2.0 by using the hand-coded decision commitment strategy (quorum threshold, for example) during testing episodes and measuring PMs 3.1 \& 3.2 throughout the episode. Thus, the Skill 2.0 curriculum would produce agents whose broadcast strategy optimises overall task performance. Finally, perhaps for future uses of ME, one does not need to strictly follow all nine steps of the Dick and Carey model \cite{dick2005systematic}. Instead, take inspiration from it, follow the model where needed and deviate where necessary.
\par
As discussed in previous literature, inherent to the collective perception problem is a speed-accuracy trade-off \cite{valentini2015efficient, valentini2016collective, hussein2021machine}. This trade-off is clear in Figures \ref{accuracyFinalChart} \& \ref{timeFinalChart}, where accuracy comes at a cost to speed. Our algorithms offer potential mechanisms to control this balance of speed and accuracy. The first potential mechanism is the quorum convergence threshold $\theta$. We hypothesise that increasing $\theta$ would decrease task time, though it would likely decrease accuracy, and vice versa for decreasing $\theta$. Another mechanism to control the balance of priorities is the ratio of the environment agents must explore before they can choose between sensing, broadcasting or committing to their opinions. For this research, we set this ratio to be $\frac{1}{L}$, where $L$ is the length of the environment. We conjecture that increasing this exploration ratio would increase accuracy at the cost of increased task time and vice versa. Figures \ref{accuracyFinalChart} \& \ref{timeFinalChart} also show that the max opinion weight $w_{max}$ can control speed and accuracy. Our results demonstrate that increasing $w_{max}$ results in quicker decisions but at a cost to accuracy, and vice versa for decreasing $w_{max}$.
\par
\begin{figure}[htbp]
\centering
    \includegraphics[width=0.8\columnwidth]{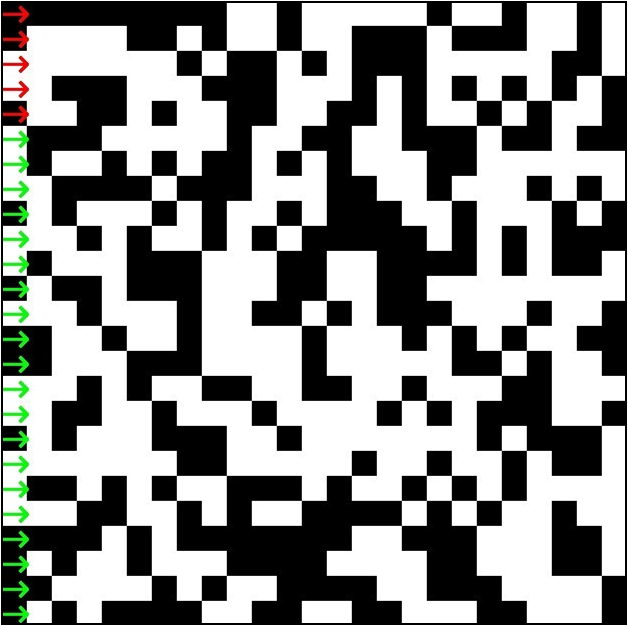}
    \caption{Environment with malicious (red arrows) and regular members (green arrows). Note the clustered distribution when initially placing malicious and regular members}
\label{distoOfMalRegularMembers}
\end{figure}

It is known that the initial spatial distribution of swarm members has a considerable effect on performance \cite{hussein2021machine}. Within our experiments, malicious and regular swarm members were initially spatially distributed in a clustered fashion, as highlighted in Figure \ref{distoOfMalRegularMembers}. Though they occupied the same grid, malicious agents were placed next to other malicious agents and regular agents were placed next to other regular agents. Thus, the only initial interaction between malicious and regular agents occurred at the border of these two clusters. If task time were extended, a sufficient mixing of agents would occur. However, task time was minimal, so the groups had limited interaction. Further, since agents can only broadcast locally, we hypothesise that the malicious influence is further limited. Future work could investigate and categorise spatial distribution into clustered and non-clustered swarms. The non-clustered distribution of malicious members would likely increase the negative influence and significantly affect task performance. Future lesson design could incorporate these spatial distributions. A non-clustered distribution of malicious members may also result in a statistically significant performance difference between opinion weighting methods.
\par
It is noted that only Skill 2.0 used ML for behaviour selection, though the \textit{parameterised probability} did achieve comparable results. These comparable results suggest that perhaps there exists a hand-coded rule that optimally dictates choosing to sense or broadcast. However, using ME education approaches with ML still has efficacy for future work. We argue this as the process of designing states and rewards for the RL algorithms brings a focus to the essential factors for a given behaviour. This focused view may ease the task of designing optimal rules for behaviour selection. Observing agents after learning a module could also educate and inspire domain experts. For example, we designed the navigation rules-based algorithm watching agents performing navigation after learning Skill 1.2. This process could be extended to all skills, allowing hand-coded rules to be designed for all behaviours. As stated throughout Section \ref{Results}, we prefer rules-based behaviour selection. Often, the rules-based algorithms require less computing resources than an ML model would require. Thus, hand-coded rules are more applicable to real-world agents who are likely limited in computing power. We are mindful of computing limits as we do not want to create infeasible solutions for the real world, as was seen for the blockchain solution \cite{strobel2018managing}.
\par
Finally, we note that our simulation is more discrete compared to previous collective perception problem simulations. Though we used the same parameters as previous experiments, the other state-of-the-art algorithms should be implemented within our simulation for a complete comparison and verification of our results.
\section{Broader Impact \label{Broader_Impact}}
Benefits from this research may include techniques and algorithms capable of developing swarms of robots resilient to environmental noise, other malfunctioning members and negative influence. Once deployed in real-world situations, given their resilience, one can trust these swarms to perform high-risk tasks or the collective consensus of the swarm. However, the insights gained through this research could have negative impacts as nefarious actors could train malicious members against these resilient swarms, thereby creating potentially super-malicious agents. However, we argue that the benefits of thinking about and developing swarm resilience techniques outweighs the potential drawbacks. For this reason, we support continued and open research in this area.
\section{Conclusions \label{Conclusions}}
Our research examined improvements to swarm resilience. Specifically, we created a simulation of the collective perception problem more agnostic to physical agent implementation and domain of operation. Machine Education approaches guided the design and creation of a curriculum that taught agents the skills needed to excel in our collective perception problem simulation. We extended previous work by implementing a skill that allowed agents to set the influence of received opinions dynamically. Our results indicated that rules-based algorithms could outperform their reinforcement learning trained counterparts. We found malicious agents can affect swarm performance by slowing task time and decreasing accuracy. However, we found a non-significant difference in performance when comparing static and dynamic weighting methods in hostile environments. Future research could develop techniques and algorithms that improve resilience further, leading to better task speed and accuracy, even in the presence of malicious members. This paper has proposed potential focuses for such future research and development. Continued study of swarm resilience and machine education techniques will identify more resilient and universal algorithms, leading to a future where swarm solutions are applied to a broader array of challenging real-world problems.
\section{Future Work \label{Future_Work}}
We suggest a few directions in which future research could head. One such direction is developing more intelligent malicious agents. These agents, for example, could seek to slow down regular agent navigation. For example, intelligent-malicious agents could seek to corner regular agents and minimise the number of tiles explored by regular agents. Cornering agents would significantly disrupt exploration and opinion diffusion. One could use RL techniques to train these intelligent-malicious agents to act in these disruptive ways. Intelligent-malicious agents could also learn to hide their identity so that regular swarm members could not detect their presence. To disguise their presence, intelligent-malicious agents could learn a sophisticated strategy that maximises disruption whilst minimising the agent's detectability. One could then develop an adversarial learning situation in which regular and malicious members are learning their subsequent goals at the same time.
\par
Another branch of research could seek to understand swarm resilience in our simulation further. As discussed in Section \ref{Discussion}, other state-of-the-art approaches should be implemented within our simulation to enable full and comprehensive comparisons. Investigations could also seek to understand the effects of:
\begin{enumerate}
    \item communication, sensing and malicious noise (both separately and combined) 
    \item swarm size on swarm resilience
    \item initial spatial distribution of the swarm on swarm resilience
    \item initial spatial distribution of malicious members amongst regular members.
\end{enumerate}
\par
Future research could also develop more swarm resilience techniques. Ideas to improve swarm resilience include:
\begin{enumerate}
    \item during training, setting the standard deviation for the truncated distribution of feature ratios inversely proportional to task performance. Consequently, a good performance would result in a tighter curved more closely centred around 0.5, and vice versa for poor performance  
    \item combining traditional approaches \cite{valentini2016collective} with the momentum-based opinion fusion
    \item using communication, sensing and malicious noise throughout all skill training modules
    \item looking to the field of information propagation \cite{zou2010fast, chen2013information, wang2015maximizing, cho2019uncertainty} for techniques and methods to improve and construct robust opinion diffusion methods.
\end{enumerate}
\par
Finally, the rules and algorithms developed in this paper should be implemented and tested within physical agents to assess and prove their applicability to real-world agents.
\bibliographystyle{plain}
\bibliography{references}
\end{document}